\documentclass[dvipsnames,runningheads]{llncs}
\usepackage{graphicx}
\usepackage{hyperref}
\usepackage{amsmath}
\usepackage{amssymb}
\usepackage{bm}
\usepackage{bbm}
\usepackage{mathtools}
\usepackage{enumitem}
\usepackage{mathpartir}
\usepackage{subfigure}
\usepackage{listings}
\usepackage{xcolor}
\usepackage{xspace}
\usepackage{tikz}
\usepackage{url}
\usepackage{tkz-euclide}
\usepackage[T1]{fontenc}

\newcommand{\Comment}[1]{}
\newcommand{\Space}[1]{}

\newcommand{\term}[1]{\emph{#1}}

\newcommand{\evaluates}{\Downarrow}

\newcommand{\R}{\mathbb{R}}

\newcommand{\N}{\mathbb{N}}

\newcommand{\proves}{\vdash}

\newcommand{\dom}{\text{dom}~}

\newcommand{\proj}{\text{proj}}

\newcommand{\implicits}{\texttt{implicit}}
\newcommand{\params}{\texttt{params}}
\newcommand{\nonvar}{\texttt{nonvar}}

\newcommand{\tand}{\ensuremath{~\text{and}~}}

\newcommand{\tif}{\ensuremath{~\text{if}~}}

\newcommand{\owise}{\ensuremath{~\text{otherwise}~}}

\newcommand{\brackets}[3]{\ensuremath{{\left#1 {#3} \right#2}}}
\newcommand{\parens}[1]{\brackets{(}{)}{#1}}

\newcommand{\bnfdef}{\ensuremath{\Coloneqq}}
\newcommand{\bnfalt}{\ensuremath{\mid}\xspace}

\newcommand{\prop}[1]{#1 ~ \text{prop}}
\newcommand{\typ}[1]{\text{typ}\parens{#1}}

\lstdefinelanguage{pecan}{
	keywords=[1]{forall, exists, max, min, sup, inf, are, is, if, then, match, with, case, end, let, be, in, else, iff},
	keywordstyle=[1]\color{blue}\bfseries,
	keywords=[2]{false, true, sometimes},
	commentstyle=\color{CadetBlue}\textit,
	stringstyle=\color{ForestGreen}, 
	keywordstyle=[2]\color{orange}\bfseries,
	keywords=[3]{assert_prop,Structure,defining,Theorem,Prove,Example,Alias,Restrict,Define,Display,Execute,load,shuffle,import,save_aut,save_aut_img,that,context,end_context,forget,shuffle,shuffle_or,using,of},
	keywordstyle=[3]\color{teal}\bfseries,
	keywords=[4]{@annotation,@postprocess,@no_simplify,@simplify,@simplify_states,@simplify_edges},
	keywordstyle=[4]\color{purple}\bfseries,
	literate=%
	    {\#}{{{\color{teal}\bfseries\#}}}1
	    {+}{{{\color{red}+~}}}1
	    {-}{{{\color{red}-~}}}1
        {:=}{{{\color{red}:=~}}}1
        {..}{{{\color{red}..~}}}1
        {\{}{{{\color{red}\{}}}1
        {\}}{{{\color{red}\}}}}1
        {|}{{{$\color{red} \lor~$}}}1
        {*}{{{\color{red}*~}}}1
        {:}{{{\color{red}:~}}}1
        {>}{{{\color{red}>~}}}1
        {<}{{{\color{red}<~}}}1
        {<=>}{{{$\color{red}\Leftrightarrow~$}}}1
        {.}{{{\color{red}.~}}}1
        {&}{{{$\color{red} \land~$}}}1
        {!}{{{$\color{red}\lnot~$}}}1
        {!=}{{{$\color{red} \neq$}}}1
        {=}{{{\color{red}=~}}}1
        {exists }{{{$\color{red}\exists$}}}1
        {forall }{{{$\color{red}\forall$}}}1,
    sensitive=false, 
    morecomment=[l]{//}, 
    morecomment=[s]{/*}{*/}, 
    morestring=[b]", 
    showstringspaces=false
}

\lstnewenvironment{pecan}
  {
    \lstset{
        language=pecan, 
        basicstyle=\small\ttfamily, 
        mathescape=true
        }
  }
  {
  }

\lstnewenvironment{pecan_output}
  {
    \lstset{
        basicstyle=\small\ttfamily,
        mathescape=true
        }
  }
  {
  }

\newcommand{\pecaninline}[1]{\lstinline[language=pecan,basicstyle=\small\ttfamily,mathescape]{#1}}

\begin{document}

\title{Pecan: An Automated Theorem Prover for Automatic Sequences using B\"uchi Automata}

\titlerunning{Pecan: An Automated Theorem Prover}

\author{Reed Oei \and
Dun Ma \and
Christian Schulz \and
Philipp Hieronymi}

\institute{University of Illinois at Urbana-Champaign, Urbana, USA\\
\email{reedoei2,dunma2,cschulz3,phierony@illinois.edu}}

\maketitle

\begin{abstract}
Pecan is an automated theorem prover for reasoning about properties of Sturmian words, an important object in the field of combinatorics on words. It is capable of efficiently proving non-trivial mathematical theorems about all Sturmian words.
\end{abstract}

\section{Introduction}

\textbf{Pecan} is a system for \emph{automated theorem proving} originally designed to decide mathematical statements about families of infinite words, in particular about Sturmian words, and based on well-known decision procedures for B\"uchi automata due to B\"uchi \cite{Buechi}. Pecan is inspired by Walnut~\cite{walnut} by Mousavi, another automated theorem prover for deciding combinatorical properties of automatic words. \textbf{Automatic words} are sequences of terms characterized by finite automata. The main motivation to create this new tool is to decide whether a statement is true for every element of an infinite family of words rather than just determining the truth of the statement for a single given words. In such a situation not every word in this family of words is automatic, but the whole family can be recognized by an automaton.
Since the infinite families of words we want to consider are often indexed by real numbers, it is convenient to work with B\"uchi automata instead of finite automata. The canonical example of such a automatic family of words are the \textbf{Sturmian words}, that is the family $(\mathbf{w}_{\alpha,\rho})$ of all words $w=(w_n)$ over the alphabet $\{0,1\}$ such that there is $\rho \in [ 0,1 )$, called the \emph{intercept}, and an irrational $\alpha \in (0,1)$, called the \emph{slope}, with
\[
w_{n}=\lfloor n\alpha +\rho\rfloor -\lfloor (n-1)\alpha +\rho\rfloor
\]
for all $n\in \N$. Using Pecan, we can automatically reprove classical and recent theorems about Sturmian words, like the fact that they are not periodic, within minutes, and even have been able to prove completely new mathematical theorems using this software. 

The idea of using automata-based decision procedures to prove theorems in combinatorics on words has been championed by Jeffrey Shallit and successfully implemented in several papers of Shallit and his many co-authors (see Shallit \cite{Shallit-survey} for a survey and Baranwal, Schaeffer, Shallit \cite{BARANWAL2021} for implementations of decision procedure for individual Sturmian words). The development of Pecan is our contribution to this exciting research program. We leave the detailed discussion of the mathematical background such as why Sturmian words can represented using automata and which statements about Sturmian words can be proved using Pecan, to the upcoming paper \cite{DecStuWor}. Here we describe the implementation of Pecan and discuss its performance.


\begin{subsection}{Related work} 
Pecan improves on Walnut~\cite{walnut}, a similar automata-based theorem prover for automatic sequences, by using B\"uchi automata instead of finite automata.
This difference enables Pecan to handle uncountable families of sequences, allowing us quantify over all Sturmian words. Additionally, the Pecan language is able to use multiple numeration systems at a time, has a concept of types outside of numeration systems, and has meta-programming language, Praline.

Many other theorem provers exist, such as SMT solvers and proof assistants, like Coq~\cite{the_coq_development_team_2020_3744225} or Isabelle~\cite{nipkow2002isabelle}.
To our knowledge, no SMT solver supports reasoning about Sturmian words.
Systems like Coq or Isabelle have projects attempting to formalize some aspects of combinatorics on words and automatic sequences~\cite{hivert2018littlewoodrichardson,holub2020binary}.
However, proofs in these systems are mostly human written, with some help from heuristics or specialized solvers, rather than being fully automatic, as in Pecan.

B\"uchi automata have also been used extensively in program verification in systems such as SPIN~\cite{gerard2003spin}.
However, we are interested in proving mathematical results, rather than proofs about properties of programs.
For this reason, we must allow unrestricted use of logical operations, such as negation, rather than restricting to more limited forms of expressing properties, such as linear temporal logic, which such systems tend to use for performance reasons.

\end{subsection}

\begin{subsection}{Acknowledgements}
Support for this project was provided by the Illinois Geometry Lab. This project was partially supported by NSF grant DMS-1654725. \end{subsection}
\section{Background}\label{sec:background}

This section contains an informal introduction to words, automata, and the notation that we use.
For precise statements and proof, we refer the reader to Allouche and Shallit \cite{MR1997038} or Khoussainov and Nerode ~\cite{aut_theory}.

Let $\Sigma^*$ denote the set of finite words on the alphabet $\Sigma$, let $\Sigma^+$ denote the set of nonempty finite words on the alphabet $\Sigma$, and let $\Sigma^\omega$ denote the set of $\omega$-words on the alphabet $\Sigma$.

For a word $w$, let $w[i]$ denote the $i$-letter of $w$.
Let $w(i,n)$ denote the length-$n$ factor of $w$ starting at $i$ and ending at $i + n - 1$, that is, $w[i \ldots i + n - 1] = w[i] w[i + 1] \cdots w[i + n - 1]$.
Let $|w|$ denote the \term{length} of $w$.

\term{B\"uchi automata} are an extension of the standard finite automata to infinite inputs.
A B\"uchi automata $\mathcal{A} = (Q, \Sigma, \delta, q_0, F)$ accepts an infinite word $w \in \Sigma^\omega$ if the run of the automaton on the word $w$ visits an accepting state (i.e., a state in $F$) infinitely many times.
The set of words accepted by $\mathcal{A}$ is its \term{language}, $L(\mathcal{A})$.
Notably, nondeterministic B\"uchi automata and \textbf{not} equivalent to deterministic B\"uchi automata, and many interesting properties are only expressible via nondeterministic B\"uchi automata.
For that reason, we simply refer to nondeterministic B\"uchi automata as B\"uchi automata, without qualification.
Additionally, when we say ``automata'' without qualification, we refer to B\"uchi automata.
Importantly, the languages that B\"uchi automata define are closed under intersection, union, projection, and complementation, and emptiness checking is decidable.

\section{Overview}\label{sec:features}

For full documentation on the features of Pecan, see the more comprehensive manual available at our repository~\cite{pecan-repo}.




\paragraph{Directives} are the interface to Pecan, instructing it to perform actions (e.g., prove a theorem).
We discuss the most important: \pecaninline{Restrict}, \pecaninline{Structure}, and \pecaninline{Theorem}.

\begin{pecan}
Restrict VARIABLES are TYPE_PREDICATE.
\end{pecan}

In all following code in the file in which the \pecaninline{Restrict} appears, the variables specified are now consider to be of the specified type.

\vspace{-0.5em}
\begin{pecan}
Structure TYPE_PREDICATE defining { FUNCTION_PREDICATES }
\end{pecan}
\vspace{-0.5em}

Defines a new structure.
The \pecaninline{TYPE_PREDICATE} is essentially the part written after the \pecaninline{is} in a restriction. 
For example, in \pecaninline{Restrict x is nat.}, the type predicate is \pecaninline{nat}; in \pecaninline{Restrict i is ostrowski(a).}, the type predicate is \pecaninline{ostrowski(a)}.
The function predicates become available to be called using the names in quotes---this feature allows for ad-hoc polymorphism, as described in Section~\ref{sec:implementation}.
It is also used to resolve arithmetic operators, such as \pecaninline{+} (which calls the relevant \pecaninline{adder}) and \pecaninline{<} (which calls the relevant \pecaninline{less}).

\vspace{-0.5em}
\begin{pecan}
Theorem ("THEOREM NAME", { PREDICATE }).
\end{pecan}
\vspace{-0.5em}
\pecaninline{Theorem} is the interface to the theorem proving capabilities of Pecan, stating that Pecan show attempt to prove the specified \pecaninline{PREDICATE} is true.

Below is an example of using all three features from above: specifying a structure called \pecaninline{nat}, restricting variables, and then proving a theorem, which is true because of the dynamic call resolution.
\vspace{-0.5em}
\begin{pecan}
Structure nat defining {
    "adder": bin_add(any, any, any),
    "less": bin_less(any, any)
}
Restrict a, b are nat.
Theorem ("", { forall a,b. a < b <=> bin_less(a,b)}).
\end{pecan}
\vspace{-0.5em}

\paragraph{Automatic Words}
Any predicate $P$ can be interpreted as a word by writing $P[i]$, which is treated as $1$ if $P(i)$ is true, and $0$ if $P(i)$ is false.
Currently only binary automatic words are supported.
We use the following translations into the IR:

\begin{itemize}
    \item $P[i] = 0 \leadsto \lnot P(i)$
    \item $P[i] = 1 \leadsto P(i)$
    \item $P[i] = Q[j] \leadsto P(i) \iff P(j)$
    \item $P[i] \neq Q[j] \leadsto \lnot (P(i) \iff P(j))$
    \item $P[i..j] = P[k..\ell] \leadsto j + k = i + \ell \land \forall n \in \typ{i}. i + n < j \Rightarrow P[i + n] = P[k + n]$
\end{itemize}

\section{Implementation}\label{sec:implementation}

This section describes the high-level the implementation of Pecan.
We give a formal definition of the Pecan language, starting with the typing rules and associated definitions in Section~\ref{sec:typing}, and then the rules for evaluation in Section~\ref{sec:evaluation-rules}.




Figure~\ref{fig:syntax} is the syntax for the core of the Pecan language.
Pecan also supports some simple syntactic sugar, such as \pecaninline{if P then Q}, which expands into \pecaninline{!P |$~$Q}, or \pecaninline{n*x} for some literal number \pecaninline{n}, which expands into \pecaninline{x+x+x+...+x} with \pecaninline{n} repetitions.
\begin{figure}
    \centering
    \vspace{-4em}
    \begin{align*}
        P,f &\in \textsc{PredicateNames} & V &\in \textsc{VariableMaps} & a,x,y,z &\in \textsc{Identifiers} \\ 
        n,m &\in \N & \mathfrak{A} &\in \textsc{Automata} \\
    \end{align*}
    \vspace{-1em}
    \begin{tabular}{l r l}
         Prog & \bnfdef & $\overline{\text{Definition}}$ \\
         $\tau$ & \bnfdef & $P$ \bnfalt $P(\overline{x})$ \\
         Definition & \bnfdef & $P(\overline{x : \tau})$ := Pred \\
         & \bnfalt & Restrict $\overline{x}$ are $P(\overline{x})$ \\
         Pred & \bnfdef & true \bnfalt false \bnfalt Pred $\lor$ Pred \bnfalt $\lnot$ Pred \bnfalt Pred $\land$ Pred \\
         & \bnfalt & $\exists x. $ Pred \bnfalt E $<$ E \bnfalt E $=$ E \bnfalt $P(\overline{E})$ \bnfalt $\text{Aut}(V, \mathfrak{A})$ \\
         E & \bnfdef & E + E \bnfalt E - E \bnfalt $x$ \bnfalt $n$ \bnfalt $f(\overline{E})$ 
    \end{tabular}
    \caption{Syntax of the core of Pecan.}
    \vspace{-1em}
    \label{fig:syntax}
\end{figure}

\subsection{Type Checking}\label{sec:typing}

A \term{type} in Pecan is represented by a B\"uchi automaton.
We say that $x : \tau$ when $x \in L(\tau)$, sometimes simply written, as an analogy to logical predicates, as $\tau(x)$.
Types may be \term{partially applied}.
For example, if $\tau = P(x_1, \ldots, x_n)$, where $P$ is some B\"uchi automaton, then $y : \tau$ when $(x_1, \ldots, x_n, y) \in L(P)$; and $\tau(y)$ holds when $y : \tau$.
In the concrete syntax of Pecan, we write \pecaninline{y is tau} or \pecaninline{y} $\in$ \pecaninline{tau}; for one or more variables, we can write \pecaninline{x, y, z are tau} to mean $x : \tau$, $y : \tau$, $z : \tau$.

The judgement $\Gamma \proves x : \tau$ means that we can prove $\tau(x)$ is true in the type environment $\Gamma$, consisting of pairs $x : \tau$.
We write the \term{domain} of $\Gamma$ as $\dom{\Gamma}$.
The judgement $\Gamma \proves \prop{P}$ means that $P$ is a \emph{well-formed proposition} in the environment $\Gamma$.
A predicate $P(\overline{x : \tau}) := Q$ is \emph{well-formed} when $\overline{x : \tau} \proves \prop{Q}$.
Below, we assume that the set of all well-formed predicates, which have already been checked, is ambiently available as $\mathcal{P}$.

\paragraph{Structures.}
In order to support ad-hoc polymorphism, Pecan allows the definition and use of \term{structures}.
This feature facilitates the use of nicer syntax for arithmetic expressions (e.g., $x + (y + z) = w$ instead of $\exists t. \texttt{adder}(x, y, t) \land \texttt{adder}(t, z, w)$) without tying ourselves to a single numeration system.
For example, $\texttt{adder}$ will be resolved to some concrete predicate predicate based on the type of $x$, $y$, and $z$.
We assume structure definitions are ambiently available in the program.

\begin{definition}
    A \term{structure} is a pair $(t(\overline{x}), D)$ where $\overline{x}$ are identifiers and $D$ is a map of identifiers to \term{call templates} of the form $f(\overline{y})$, where for each $\overline{y} \subseteq \overline{x} \cup \{*\}$; $*$ denotes ``any.''
    The \term{name} of the structure is $t$.
\end{definition}

We write the sequence of indexes of the arguments that are $*$, called \term{parameters}, as $\params(f(\overline{y}))$.
A call template is $n$-ary if $|\params(f(\overline{y}))| = n$.
The sequence of the indexes of the other arguments (i.e., not $*$), called \term{implicits}, is $\implicits(f(\overline{y}))$.
For example, $\params(f(a, *, b, *)) = [2, 4]$ and $\implicits(f(a, *, b, *)) = [1, 3]$.
We assume that typechecking has been done before evaluating, because we may need structure information at runtime to resolve \term{dynamic calls}, that is, calls whose name matches some definition inside a structure.
We denote the type that an expression $e$ got when typechecking by $\typ{e}$.

We write $t[P] = Q(y_1, \ldots, y_m)$ to look up a definition in the associated map $D$, and we say that $t$ \term{has an $m$-ary definition} for $P$ in this case.
If $t$ does not have a definition for $P$, then we write $t[P] = \bot$.

\begin{definition}
    A structure is called \term{numeric} if it has a ternary definition for $\texttt{adder}$ and a binary definition $\texttt{less}$.
    We write $x + y = z$ when $\texttt{adder}(x, y, z)$ holds and $x < y$ when $\texttt{less}(x, y)$ holds.
    A numeric structure may also optionally contain the following definitions:
    \begin{itemize}
        \item A binary definition $\texttt{equal}(x,y)$, written $x \equiv y$.
            The default is equality, $x = y$.
        
        \item A unary definition $\texttt{zero}(z)$.
            The default is $z = 0^{\omega}$.
            
        \item A unary definition $\texttt{one}(x)$.
            The default is $0 \leq x \land \forall y. y = 0 \lor x \leq y$.
    \end{itemize}
\end{definition}


\begin{definition}
    We can \term{resolve} a call $P(\overline{e})$ as $Q(\overline{a : \tau})$, written $P(\overline{e}) \leadsto Q(\overline{a : \tau})$, if $Q(\overline{a : \tau}) \in \mathcal{P}$ and for some structure $t(x_1, \ldots, x_\ell)$, for each $1 \leq i \leq |\overline{e}|$, either:
    \begin{enumerate}
        \item $\typ{e_i} = t(x_1, \ldots, x_\ell)$, and $t[P] = Q(b_1, \ldots, b_m)$ such that
        for each $1 \leq j \leq m$,
        \[
            a_j =
            \begin{cases}
                x_k & \tif \implicits(Q(b_1, \ldots, b_m))[k] = j \\
                e_k & \tif \params(Q(b_1, \ldots, b_m))[k] = j
            \end{cases}
        \]
        
        \item $\typ{e_i} = s(y_1, \ldots, y_p)$, where $s \neq t$ and $s[P] = \bot$.
    \end{enumerate}
    
    or, if none of the arguments have a definition for $P$, then $P(\overline{e}) \leadsto P(\overline{a : \tau})$.
\end{definition}

\framebox{$\Gamma \proves e : \tau$} \textbf{Expression Typing}
The expression typing rules are standard, with the exception that we can treat any predicate as a function by not writing it's last argument (e.g., $f(x)$ denotes the $y$ such that $f(x,y)$ holds).
\begin{mathpar}
\small
\inferrule*[right=Int]{
    \tau~\text{is numeric}
} { \Gamma \proves i : \tau }

\inferrule*[right=Var]{
    x : \tau \in \Gamma
} { \Gamma \proves x : \tau }

\inferrule*[right=Op]{ 
    \Gamma \proves a : \tau
    \\ 
    \Gamma \proves b : \tau
    \\
    \oplus \in \{ +, - \}
}{ \Gamma \proves a \oplus b : \tau }

\inferrule*[right=Func]{
    \Gamma \proves \overline{e : \tau}
    \\
    (f(\overline{x : \tau}, r : \sigma) := e') \in \mathcal{P}
}{ \Gamma \proves f(\overline{e}) : \sigma }
\end{mathpar}

\framebox{$\Gamma \proves \prop{P}$} \textbf{Well-formed Propositions}
\begin{mathpar}
\small
\inferrule*[right=Rel]{ 
    \Gamma \proves a : \tau
    \\
    \Gamma \proves b : \tau
    \\
    \Join \in \{ \equiv, < \}
}{ \Gamma \proves \prop{a \Join b} }

\inferrule*[right=Aut]{
}{ \Gamma \proves \prop{\text{Aut}(V, \mathcal{A})} }

\inferrule*[right=BinPred]{
    \Gamma \proves \prop{P}
    \\
    \Gamma \proves \prop{Q}
    \\
    \oplus \in \{ \lor, \land \}
}{ \Gamma \proves \prop{P \oplus Q} }

\inferrule*[right=Comp]{
    \Gamma \proves \prop{P}
}{ \Gamma \proves \prop{\lnot P} }

\inferrule*[right=Exists]{
    \Gamma, x : \tau \proves \prop{P}
}{ \Gamma \proves \exists x : \tau. \prop{P} }

\inferrule*[right=Call]{
    \Gamma \proves \overline{e : \tau}
    \\
    (f(\overline{x : \tau}) := e') \in \mathcal{P}
}{ \Gamma \proves \prop{f(\overline{e})} }

\end{mathpar}

\subsection{Evaluation}\label{sec:evaluation-rules}

Pecan is a simple tree-walking interpreter written in Python 3~\cite{python3} which typechecks and processes each top-level construct in order.
Operations with non-trivial implementations are described in detail below.
Most basic automata operations (e.g., conjunction, disjunction, complementation, emptiness checking, simplification) are implemented using the Spot library \cite{duret.16.atva2}. 

\textbf{Automata Representation}
Automata are represented by a pair of $(V, \mathcal{A})$, where $V$ is a map taking variable names to an ordered list of APs that represent it, called the \emph{variable map}, and $\mathcal{A}$ is a Spot automaton (specifically, a value of type \texttt{spot.twa\_graph}).
We use the convention that calligraphic letters represent actual B\"uchi automata, and Fraktur letters represent automata in the Pecan sense of a pair of a variable map and B\"uchi automaton.

A \emph{variable map} $V$ is a finite set of mappings $x \mapsto [\texttt{ap}_1, \ldots, \texttt{ap}_n]$ such that for all distinct variables $x$ and $y$, $V[x] \cap V[y] = []$.
We denote by $V[x]$ the list of APs that $x$ is represented by, and we denote by $V \cup W$ the union of two variables maps union, which is only defined when the only keys that $V$ and $W$ have in common have identical APs.
$V \sqcup W$ is the disjoint union of these maps.
$V \setminus K$ is the variable map containing every entry $x \mapsto a \in V$ such that $x \not\in K$.

For two variable maps $V$ and $W$, $V \ll W$ denotes their \emph{biased merge}, which is a pair $(U, \theta)$ of a variable map $U$ and a substitution $\theta$ such that $U = V \cup W\theta$.
A substitution is a set of mappings $a \mapsto b$ where $a$ and $b$ are both APs, which can be applied to a variable map or an automaton to rename the APs in them.
For example, if $\theta = \{ a \mapsto d, c \mapsto e \}$, then $\{ x \mapsto [ a, b, c ] \} \theta = \{ x \mapsto [ d, b, e ] \}$.
When it is clear, we also write $V \ll W$ to denote just the resulting variable map, without the associated substitution.

Below, we describe the evaluation of Pecan programs via a big-step relation $E \evaluates \mathcal{A}$.
Automata literals (generally loaded from files), written $\text{Aut}(V, \mathcal{A})$, simply evaluate to be the automata they store: $\text{Aut}(V, \mathcal{A}) \evaluates (V, \mathcal{A})$.

\textbf{Logical Operations}
Fundamental automata operations (i.e., $\land$ and $\lor$, represented by $\oplus$ below) are defined below.
\[
\small
    (V, \mathcal{A}) \oplus (W, \mathcal{B}) = 
    \begin{cases}
        (V \ll W, \mathcal{A} \oplus \mathcal{B}) & \tif |S(\mathcal{A})| < |S(\mathcal{B})| \\
        (W \ll V, \mathcal{A} \oplus \mathcal{B}) & \owise
    \end{cases}
\]
where $S(\mathcal{A})$ denotes the set of states of $\mathcal{A}$.
We also define $\lnot (V, \mathcal{A}) = (V, \lnot \mathcal{A})$.

\textbf{Substitution}
Let $\mathfrak{A} = (V, \mathcal{A})$, where $\mathcal{A} = (Q, \Delta, \delta, q_0, F)$ be a B\"uchi automaton where $\Delta$ is the set of formulas involving $\land$, $\lor$, and $\lnot$ on a finite set $X$ of APs.
We now define the substitution $\mathcal{A}[y/x]$, replacing $x$ by $y$.
Let $A = [\overline{a}]$ be the list of APs representing $x$ (i.e., $A = V[x]$), and let $B = [\overline{b}]$ be the list of APs representing $y$, which we assume is ambiently available.
This can be stored globally, and generated when needed if the variable $y$ has never been used before.

Define $\mathfrak{A}[y/x] = (V', \mathcal{A}')$ where $V' = (V \setminus \{ x \}) \cup \{ y \mapsto B \}$, and $\mathcal{A}' = (Q, \Delta', \delta', q_0, F)$, with the new set of variables $X' = (X \setminus A) \cup B$ and the same underlying alphabet, such that:
\[
    \Delta' = \{ \varphi[\overline{b/a}] : \varphi \in \Delta \}; \tand \delta' = \{ (s, d, \varphi[\overline{b/a}]) : (s, d, \varphi) \in \delta' \}
\]

\textbf{Predicate Calls}
\begin{mathpar}
\small
\inferrule*[right=Call]{
    \overline{e \evaluates (\mathfrak{A}, x)}
    \and
    \nonvar(\overline{e}) = [k_1, \ldots, k_{\ell}]
    \\
    P(\overline{X}) \leadsto Q(\overline{y})
    \and
    (Q(\overline{z : \tau}) := R) \in \mathcal{P}
    \and
    R \evaluates \mathfrak{B}
}{ P(\overline{e}) \evaluates \proj_{x_{k_1}, \ldots, x_{k_{\ell}}}\left( \bigwedge \overline{\mathfrak{A}} \land \mathfrak{B}[\overline{y/z}]\right) }
\end{mathpar}
where $\nonvar(\overline{e})$ denotes the nonvariable positions in $\overline{e}$.



\textbf{Existential Quantification}
\begin{mathpar}
\small
\inferrule*[right=Exist]{
    (\tau(x) \land P) \evaluates (V, \mathcal{A})
}{ (\exists x \in \tau. P) \evaluates \proj_{V[x]}(V, \mathcal{A}) }
\end{mathpar}

Here $\proj_{V[x]}(\mathfrak{A})$ denotes the automaton $\mathfrak{A}$ after projecting out every AP representing $x$ in the variable map $V[x]$; this operation is implemented in Spot.

\textbf{Expressions}
An expression $E$ evaluates to a pair $(\mathfrak{A}, x)$ of an automaton $\mathfrak{A}$ and a variable $x$.
Many rules, like \textsc{Add}, need to evaluate subexpressions.
While evaluating a subexpression $e$, it may be that we generate fresh variables to store the result, which must be projected out.
The only case in which this does not occur is when the subexpression is itself a variable.
We write $\proj_{\overline{x}}(\mathfrak{A})$ to denote projecting out the intermediate variables resulting from computing expressions that are \textbf{not} variables.
For example, if $a \evaluates (\mathfrak{A}, x)$ and $b \evaluates (\mathfrak{B}, y)$ then $\proj_{a,b}(\mathfrak{A})$ denotes $\proj_V(\mathfrak{A})$ where $V = \{ v : (e, v) \in \{ (a,x), (b,y) \}, e \neq v \}$.

\begin{mathpar}
\small
\inferrule*[right=Var]{ }{ x \evaluates (\top, x) }

\inferrule*[right=Zero]{ x ~ \text{fresh} }{ 0 \evaluates (\texttt{zero}(x), x) }

\inferrule*[right=One]{ x ~ \text{fresh} }{ 1 \evaluates (\texttt{one}(x), x) }

\inferrule*[right=Add]{ 
    a \evaluates (\mathfrak{A}, x)
    \\
    b \evaluates (\mathfrak{B}, y)
    \\
    (x + y = z) \evaluates \mathfrak{C}
    \\
    z ~ \text{fresh}
} { a + b \evaluates (\proj_{a,b}(\mathfrak{A} \land \mathfrak{B} \land \mathfrak{C}), z) }

\inferrule*[right=Sub]{ 
    a \evaluates (\mathfrak{A}, x)
    \\
    b \evaluates (\mathfrak{B}, y)
    \\
    (z + y = x) \evaluates \mathfrak{C}
    \\
    z ~ \text{fresh}
} { a - b \evaluates (\proj_{a,b}(\mathfrak{A} \land \mathfrak{B} \land \mathfrak{C}), z) }

\inferrule*[right=Int]{
    \overbrace{1 + 1 + \cdots + 1}^{n~\text{times}} \evaluates (\mathfrak{A}, x)
}{ n \evaluates (\mathfrak{A}, x) }

\inferrule*[right=Func]{
    f(\overline{e}, x) \evaluates \mathfrak{A}
    \\ x ~ \text{fresh}
}{ f(\overline{e}) \evaluates (\proj_{\overline{e}}(\mathfrak{A}), x) }

\inferrule*[right=Rel]{ 
    a \evaluates (\mathfrak{A}, x)
    \and
    b \evaluates (\mathfrak{B}, y)
    \\
    (x \Join y) \evaluates \mathfrak{C}
    \\
    \Join \in \{ \equiv, < \}
} { a \Join b \evaluates \proj_{a,b}(\mathfrak{A} \land \mathfrak{B} \land \mathfrak{C}) }
\end{mathpar}





\section{Evaluation}\label{sec:evaluation}

We evaluate the performance of Pecan by generating automata for fundamental definitions in the field of combinatorics on words and proving theorems about Sturmian words using these definitions.
We consider \term{characteristic} Sturmian words, i.e., where the intercept is $0$, which we write $c_{\alpha} = \mathbf{w}_{\alpha,0}$; all definitions are parameterized by the slope of Sturmian word.
To our knowledge, there are no other tools to which Pecan can be directly compared.
Our results indicate that our approach is practical, as we are able to prove many interesting theorems using only an ordinary computer.
There is not space to discuss the definitions and theorems encoded, but our repository contains the complete code~\cite{sturmian-words-repo}.

We record several metrics for each predicate: the number of \term{atoms}, how many \term{alternating quantifier blocks} it contains (i.e., alternating universal and existential quantifiers), the runtime in seconds, the number of states and edges in the intermediate automaton with the greatest number of states, and the final number of states and edges, when applicable.
Alternating quantifier blocks increase the runtime due to the encoding of $\forall x. P(x)$ as $\lnot(\exists x. \lnot P(x))$, as complementing B\"uchi automata has a very poor worst-case complexity of at least $\Omega((0.76n)^n)$~\cite{Yan2008}.
We write these blocks as $\forall^{n_1}\exists^{n_2}\forall^{n_3}\ldots$.
Quantifiers range over countable domains unless otherwise noted; $\forall_\R$ and $\exists_\R$ are quantifiers ranging over domains of cardinality $|\R|$. 

As an example of computing these metrics, consider the following definition.
\begin{definition}
    A factor $x$ of a word $w$ is \term{special} if $x0$ and $x1$ are factors of $w$.
\end{definition}
In Pecan, we can define this for Sturmian words as follows.
\vspace{-0.5em}
\begin{pecan}
Restrict a is bco_standard.
Restrict i,j,k,n are ostrowski(a).
special_factor(a,i,n) :=
    (existsj. factor_lt_len(a,i,n,j) & $\$$C[j+n] = 0) &
    (existsk. factor_lt_len(a,i,n,k) & $\$$C[k+n] = 1)
\end{pecan}
\vspace{-0.5em}
The numeric structure \pecaninline{ostrowski(a)} specifies the numeration system for the variables \pecaninline{i}, \pecaninline{j}, \pecaninline{k} and \pecaninline{n} making Sturmian words into automatic sequences, and \lstinline[mathescape, language=pecan, basicstyle=\small\ttfamily]!$\$$C[i]! denotes the $i$-th letter of the Sturmian word---the same automaton works for every slope.
Pecan expands this to:
\vspace{-0.5em}
\begin{pecan}
special_factor(a,i,n) :=
  (existsj. ostrowski(a,j) & factor_lt_len(a,i,n,j) & 
    !(existsv0. adder(j,n,v0) & $\$$C(v0))) & 
  (existsk. ostrowski(a,k) & factor_lt_len(a,i,n,k) & 
     (existsv1. adder(k,n,v1) & $\$$C(v1)))
\end{pecan}
\vspace{-0.5em}
We can see that \pecaninline{special_factor} has $8$ atoms and has complexity $\exists^3\forall$ in prenex normal form.
Here, \pecaninline{factor_lt_len(a,i,n,j)} means $c_{a}[i..i+n] = c_{a}[j..j+n]$.

\begin{figure}
    \centering
    \hspace{-3em}
    \vspace{-1em}
    \footnotesize
    \begin{tabular}{l|l|r|r|r|r|r|r|r|}
        & & & & \multicolumn{2}{c}{Max} & \multicolumn{2}{c}{Final} \\
        Name & Complexity & Atoms & Runtime (s) & States & Edges & States & Edges \\ \hline
Mirror invariant & $\exists$ & $1$ & $8.1$ & $1440$ & $16840$ & $1129$ & $9666$ \\
Unbordered & $\exists^3$ & $2$ & $0.5$ & $275$ & $1156$ & $92$ & $410$ \\
Cube & $\exists$ & $4$ & $0.7$ & $936$ & $5956$ & $126$ & $561$ \\
Least period & $\forall$ & $4$ & $2605.2$ & $352577$ & $6098198$ & $577$ & $4161$ \\
Max unbordered subfactor & $\forall$ & $4$ & $26.4$ & $25200$ & $196575$ & $585$ & $4345$ \\
Palindrome & $\exists^2$ & $4$ & $5.1$ & $1934$ & $12337$ & $922$ & $6274$ \\
Period & $\exists^2$ & $5$ & $64.1$ & $5853$ & $103886$ & $1660$ & $17570$ \\
Recurrent & $\forall\exists$ & $5$ & $272.6$ & $61713$ & $960207$ & $34$ & $212$ \\
Special factor & $\exists^3\forall$ & $8$ & $1361.8$ & $17738$ & $103274$ & $4594$ & $25349$ \\
Factor Lt (idx) & $\exists \forall^2$ & $11$ & $702.7$ & $1057221$ & $22348882$ & $2204$ & $25026$ \\
Eventually periodic & $\exists^2\forall\exists^2$ & $12$ & $216.6$ & $78338$ & $1001075$ & $1$ & $0$ \\
Reverse factor & $\exists \forall^2$ & $12$ & $842.0$ & $1408050$ & $22780414$ & $1440$ & $16840$ \\
Antipalindrome & $\exists^2\forall^3$ & $13$ & $242.2$ & $78396$ & $1668960$ & $200$ & $834$ \\
Antisquare & $\forall^3$ & $13$ & $1844.3$ & $2542937$ & $31570114$ & $136$ & $539$ \\
Square & $\forall^3$ & $13$ & $2138.0$ & $1908657$ & $23683717$ & $155$ & $747$ \\
$(01)^*|(10)^*$ & $\forall$ & $16$ & $77.9$ & $5409$ & $72739$ & $103$ & $456$ \\
    \end{tabular}
    \vspace{-0.5em}
    \caption{Common definitions about Sturmian words.}
    \vspace{-2em}
    \label{fig:def-performance-table}
\end{figure}

Figure~\ref{fig:def-performance-table} shows performance statistics for creating the automata representing various common definitions in Pecan.
The automaton for Eventually Periodic is empty because of the classic result that there are no Sturmian words that are eventually periodic.
One might guess that Cube would be more expensive than Square; however, we can define Cube very efficiently in terms of Square.
The same is true for higher powers, as well as many other predicates: for example, both Mirror Invariant and Palindrome are relatively easy to compute, as they are straighforwardly defined using Reverse Factor.
We can see that, even though the automata often become quite large (e.g., having over $2$ million states in the case of Antisquare), we are still able to handle them relatively easily.

\begin{figure}
    \centering
    \footnotesize
    \vspace{-1em}
    \begin{tabular}{l|r|r|r|r|r|r|}
        & & & & \multicolumn{2}{c}{Avg Max} \\
        Complexity & Atoms & Number & Avg Runtime (sec.) & States & Edges \\ \hline
$\forall_\R\exists$ & $3$ & $1$ & $0.0$ & $12.0$ & $38.0$ \\
$\forall_\R\exists^2$ & $3$ & $1$ & $0.1$ & $868.0$ & $5107.0$ \\
$\exists_\R\exists^2$ & $4$ & $2$ & $0.0$ & $53.0$ & $124.0$ \\
$\forall_\R\forall$ & $4$ & $1$ & $0.1$ & $130.0$ & $516.0$ \\
$\forall_\R\exists^2$ & $5$ & $1$ & $0.1$ & $399.0$ & $2053.0$ \\
$\forall_\R\forall^2$ & $5$ & $2$ & $0.2$ & $146.5$ & $603.0$ \\
$\forall_\R\exists\forall^2$ & $6$ & $2$ & $0.2$ & $598.5$ & $3789.5$ \\
$\forall_\R\exists^2$ & $6$ & $1$ & $0.1$ & $812.0$ & $4770.0$ \\
$\forall_\R\forall\exists$ & $6$ & $1$ & $7.1$ & $1328.0$ & $8985.0$ \\
$\forall_\R\exists^3$ & $7$ & $2$ & $0.3$ & $119.5$ & $271.5$ \\
$\forall_\R\forall\exists^2$ & $7$ & $1$ & $0.0$ & $593.0$ & $3355.0$ \\
$\forall_\R\forall^2$ & $7$ & $3$ & $9.8$ & $1746.0$ & $15430.3$ \\
$\forall_\R\forall^3$ & $7$ & $1$ & $0.1$ & $155.0$ & $1497.0$ \\
$\forall_\R\forall\exists^2$ & $8$ & $1$ & $1.4$ & $922.0$ & $6274.0$ \\
$\forall_\R\forall^2$ & $9$ & $2$ & $0.1$ & $178.0$ & $848.5$ \\
$\forall_\R\forall^2\exists$ & $10$ & $1$ & $0.2$ & $1440.0$ & $16840.0$ \\
$\forall_\R\forall\exists\forall\exists\forall\exists$ & $17$ & $1$ & $3.3$ & $6106.0$ & $46025.0$ \\
$\forall_\R\forall^4\exists^4$ & $18$ & $1$ & $156.6$ & $2032240.0$ & $47851215.0$ \\
$\forall_\R\forall\exists^3\forall^2$ & $22$ & $1$ & $489.8$ & $138223.0$ & $3834628.0$ \\
    \end{tabular}
    \vspace{-0.5em}
    \caption{Theorems about Sturmian words, grouped by complexity and number of atoms.
        Number column shows how many theorems are in each group.
        Theorems evaluate to single state automata, so we omit the data about the final automaton.}
    \vspace{-2em}
    \label{fig:thm-performance-table}
\end{figure}

Figure~\ref{fig:thm-performance-table} shows performance statistics for proving theorems about Sturmian words in Pecan.
These theorems are a mix of classical results, known theorems, and some new results we proved using Pecan, described in \cite{DecStuWor}.
Overall, our results show that Pecan is a viable theorem proving tool for Sturmian words, and we hypothesize it will also be useful for other B\"uchi-automatic sequences.

\section{Conclusion and Future Work}\label{sec:conclusion}

We presented Pecan, the first system, to our knowledge, implementing a general purpose decision procedure for B\"uchi-automatic sequences, and in particular, statements about Sturmian words.
The system aims to be a convenient interface for specifying definitions and proving theorems about such sequences, with features such as custom numeration systems, enabled by the type system, and convenient syntax for indexing into automatic sequences.
We provide a formal description of the system, then evaluate the performance of Pecan by building automata representing common definitions and proving theorems about Sturmian words.
We show that Pecan has reasonable performance, despite the theoretical worst-case, indicating that the approach is practical.

In the future, we hope to expand the statements that Pecan is capable of handling, by integrating known extensions such as multiplication by quadratic irrationals, as described in~\cite{hieronymi2019presburger}.
We also hope to continue using Pecan to prove theorems about Sturmian words, both to provide new proofs of old results, as well as proving more new results.
It may also be interesting to support more expressive kinds of automata that still have the desired closure properties, such as ($\omega$-)operator precedence automata~\cite{panella2013operator}.
We would also like to integrate Pecan into a general-purpose proof assistant, such as Isabelle or Lean~\cite{de2015lean}. 

\bibliographystyle{splncs04}
\bibliography{biblio.bib}

\end{document}